\documentclass[12pt,]{article}
\usepackage{amsmath}
\usepackage{amsfonts} \usepackage{amssymb}

\numberwithin{equation}{section}

\begin{document}      

\title{
Cosmological 
Einstein-$\lambda$-perfect-fluid solutions with 
asymptotic dust or radiation equations of state.\\
}

\author {Helmut Friedrich\\
Max-Planck-Institut f\"ur Gravitationsphysik\\
Am M\"uhlenberg 1\\ 14476 Golm, Germany }

\maketitle

\begin{abstract}

This article introduces the notions of  {\it asymptotic dust} and  {\it asymptotic radiation equations of state}. With these non-linear generalizations of the well known {\it dust} or (incoherent) {\it radiation} equations of state the perfect-fluid equations loose any conformal covariance or privilege. 
We analyse the conformal field equations induced with these equations of state. It is shown that the
 Einstein-$\lambda$-perfect-fluid equations with an
 asymptotic radiation equation of state allow for large sets of data that develop into solutions which admit smooth conformal boundaries in the future and  smooth extensions beyond.

\end{abstract}

{\footnotesize


\newpage

\vspace{1cm}

\section{Introduction}

This article extends our investigation of the long term behaviour of solutions to Einstein's equation
\begin{equation}
\label{einst}
R_{\mu\nu}[\hat{g}] - \frac{1}{2}\,R[\hat{g}]\,\hat{g}_{\mu\nu}
+ \lambda\,\hat{g}_{\mu\nu} = \hat{T}_{\mu\nu},
\end{equation}
with positive cosmological constant $\lambda$
and energy-momentum tensor $ \hat{T}_{\mu\nu}$
for various matter models. The models considered in the present paper are given by  perfect fluids with certain non-linear equations of state.
Several authors studied the future asymptotic behaviour
of solutions to the Einstein-$\lambda$-perfect fluid equations
with a linear equation of state
$\hat{p} = w(\hat{\rho}) = w_*\,\hat{\rho}$, $w_* = const.$
\cite{hadzic-speck-2015}, \cite{oliynyk:2016}, 
\cite{rodnianski:speck:2013}, \cite{speck:2012}, often assuming that
$0 < w_* < 1/3$.
More general equations of state have been considered in the articles  \cite{liu:wei},  \cite{reula:1999}. In the present article the asymptotic behaviour is analysed from a particular point of view which is motivated by the following observation.

De Sitter space is a geodesically complete, spatially compact, conformally flat solution of FLRW type to the Einstein-$\lambda$-vacuum  equations with cosmological constant $\lambda  > 0$. It admits
smooth conformal boundaries at future and past time-like infinity. 
It generalizes as follows.
Smooth Cauchy data for the same equations on a Cauchy hypersurface $S$ of de Sitter space
that are sufficiently close (with respect to suitable Sobolev norms)
 to the de Sitter data
develop into solutions that are also time-like and null geodesically complete and admit  smooth conformal boundaries.
The {\it conformal Einstein equations}, which have been used for this purpose, determine  in fact
smooth solutions that extend beyond these boundaries and define  on `the other sides' again vacuum solutions. Perturbations of the conformal curvature, only restricted by the smallness condition on $S$, travel unimpeded across the boundaries \cite{friedrich:1986a}, \cite{friedrich:1986b}.

The question of interested in this article is: For which matter models can be established similar results on the future asymptotic behaviour 
of solutions to the Einstein-$\lambda$-matter equations ?
For a discussion of the answers obtained so far the reader is referred to the article \cite{friedrich:2023a}, which can also be read as an extended
 non-technical introduction to the present article. It further explains our interest in the question above and also may be helpful for readers not acquainted with the conformal methods used in the following.

The two simplest cases with a positive answer are the FLRW-type solutions to the 
Einstein-$\lambda$-perfect-fluid equations with 
the {\it pure (incoherent) radiation} equation of state 
$\hat{p} = w(\hat{\rho}) = \frac{1}{3}\,\hat{\rho}$
and the {\it pure dust} equation of state
 $\hat{p} = w(\hat{\rho}) = 0$ (the unusual word {\it pure} has been added here to avoid confusions with notions introduce below).

 These cases have been generalized by showing that these FLRW solutions are future stable in the class general solutions to the Einstein-$\lambda$-perfect-fluid equations with the respective equations of state. The perturbed solutions are not only time-like geodesically future complete but also admit smooth future conformal boundaries and extensions beyond
 \cite{friedrich:dust:2016}, \cite{luebbe:valiente-kroon:2013}. These cases are, however, still somewhat special. The perfect fluid equations with a pure radiation equation of state considered in \cite{luebbe:valiente-kroon:2013} are conformally invariant with vanishing trace of the energy momentum tensor.
The perfect fluid equations with a pure dust equation of state are not conformally invariant and have an energy momentum tensor
with non-vanishing trace. They are, however, {\it conformally privileged}
by being related in some sense to a conformally invariant structure which helps establish the result
 of  \cite{friedrich:dust:2016}.

\vspace{.1cm}

It may be reasonable to base models of the universe
on solutions to the Einstein-$\lambda$-perfect-fluid equations where  
 the future development is determined by one of the equations of state above. It appears unlikely, however, that it makes sense to
cover the whole stretch from the Big Bang to future time-like infinity 
 by one and the same linear equation of state. One can expect to need a transition where
 \[
 \hat{p} = w^{**}(\hat{\rho})\,\hat{\rho},
 \]
with some function $ w^{**}(\hat{\rho})$ of which we will assume that
$0 \le  w^{**}(\hat{\rho}) \le 1/3$. At late time $ w^{**}(\hat{\rho})$ should then acquire the value $w^{**}(\hat{\rho}) = 0$ if we want to model the end by a pure dust equation of state
and  $w^{**}(\hat{\rho}) = 0$ in the case of pure radiation.

There is nothing, however,  which fixes a natural notion of `late time'. The only meaningful requirement would be that these values are assumed in the limit when the space-time approaches future time-like infinity. But then the equation of state would still need to recognize
where and when this limit will be achieved.

In the conformal analysis of the two FLRW models mentioned above the 
physical density $\hat{\rho}$ and the conformal density $\rho$ satisfy a relation of the form
\begin{equation}
\label{rho-e-hat(rho)}
\hat{\rho} = \Omega^{e}\rho,
\end{equation}
with $e = 3$ in the case of pure dust and $e = 4$ in the case of pure radiation. These values are chosen because they imply that $\rho = \rho_* = const. > 0$ (assuming that $\hat{\rho} > 0$ on some initial slice). Since then $\hat{\rho} \rightarrow 0$ 
at future time-like infinity where $\Omega \rightarrow 0$, the behaviour of $\hat{\rho}$ can be understood as an indicator for the approach to future time-like infinity. In generalizing the situation we shall keep 
(\ref{rho-e-hat(rho)}), hoping it to serve as an indicator function of the far future where $\Omega \rightarrow 0$. The relation $\rho = \rho_* = const.$ may not be preserved under the generalization considered below and we will have to control that $\rho$ remains positive and bounded in the limit  $\Omega \rightarrow 0$. The pure dust and pure radiation equations of state are now generalized as follows.

\vspace{.1cm}

\noindent
An {\it asymptotic dust equation of state} is given by a function of the form
\begin{equation}
\label{as-dust-eos}
\hat{p} = w(\hat{\rho}) 
= \left(\hat{\rho}^{k}\,w^*(\hat{\rho})\right)\hat{\rho}
\quad \mbox{with some} \quad k \in \mathbb{N},
\end{equation}
combined with  (\ref{rho-e-hat(rho)}) where $e = 3$. It implies
$w'(\hat{\rho}) = (1 + k) \,\hat{\rho}^{k}\,\,w^*(\hat{\rho})
+ \hat{\rho}^{1 + k}\,\,(w^*)'(\hat{\rho})$, where here and in the following the notation $ ' = \partial/\partial \hat{\rho}$ is used.

\vspace{.1cm}

\noindent
An {\it asymptotic radiation equation of state} is given by a function of the form 
\begin{equation}
\label{as-rad-eos}
\hat{p} = w(\hat{\rho}) = \left(\frac{1}{3}
- \hat{\rho}^{k}\,\,w^*(\hat{\rho})\right)\hat{\rho} 
\quad \mbox{with some} \quad k \in \mathbb{N},
\end{equation}
combined with  (\ref{rho-e-hat(rho)}) where $e = 4$. It implies
$w'(\hat{\rho}) = \frac{1}{3} - (1 + k) \,\hat{\rho}^{k}\,\,w^*(\hat{\rho})
- \hat{\rho}^{1 + k}\,\,(w^*)'(\hat{\rho})$.
In both case $w^*(\hat{\rho})$ is a smooth functionof $\hat{\rho}$ that satisfies
\[
0  < w^*(\hat{\rho}) <  c = const.
\]
To ensure that $0 \le  w^{**}(\hat{\rho}) \le 1/3$ as required above, we would need to impose more detailed conditions on  
$w^*(\hat{\rho})$ but for the analysis of the effect of the new equations of state in the far future where $\Omega$ becomes small the conditions above are sufficient. The limits $w^*(\hat{\rho}) \rightarrow 0$ give back the pure dust and the pure radiation equations of state.

\vspace{.1cm}

The factor $\hat{\rho}^k$ has been inserted in the definitions 
as a simple means to control in terms of $k$ the speed at which the pure dust or the pure radiation situations is approximated as $\hat{\rho} \rightarrow 0$.
Eventually the conformal field equations may force us to impose less crude conditions on $w^*(\hat{\rho})$ or to introduce more subtle definitions of asymptotic equations of state.

\vspace{.1cm}

Definition (\ref{as-dust-eos})  implies 
 \[
\hat{T} = \hat{g}^{\mu\nu}\,T_{\mu\nu} = 3\,w(\hat{\rho}) - \hat{\rho}
= 
- \Omega^{3}\,\rho + 3\,\Omega^{3 + 3\,k}\,\rho^{1+ k}\,w^*(\Omega^{3}\,\rho),
\]
while definition (\ref{as-rad-eos}) gives
\[
\hat{T} =  - 3\,\Omega^{\,4 + 4\,k}\,\rho^{1 + k}\,w^{*}(\Omega^{4}\,\rho).
\]
In both case $\hat{T} \neq 0$ if $\rho > 0$ and  
$\hat{T} \rightarrow 0$ as $\Omega \rightarrow 0$  if $\rho$ {\it remains bounded in this limit}.
  
 Because the principal parts  of the matter equations are affected by the equations of state above any conformal covariance or privilege is lost.

\vspace{.1cm}

The Cauchy problem local in time for Einstein-$\lambda$-perfect fluids
with an asymptotic dust or radiation equations of state as above poses no problems. This  follows from the results of
\cite{friedrich98}, \cite{friedrich:rendall} where only weak conditions on the equation of state are assumed.

\vspace{.1cm}

First conditions on the admissible values of $k$ are found if the equations of state above are considered in the
conformal analogues of the FLRW Friedmann and the energy conservation equation. In the case of asymptotic dust the systems reads  
\[
(\dot{\Omega})^2 = \frac{\lambda}{3}
- \frac{R[\hat{h}]}{6}\,\Omega^2 + 
\Omega^{3}\,\frac{\rho}{3},
\quad \quad
\dot{\rho} = \Omega^{3\,k - 1}\,\rho^{1 + k}\,w^*(\Omega^3\,\rho))\,\dot{\Omega},
\]
where $R[\hat{h}] = const. \ge 0$.
It can be integrated across $\Omega = 0$ with $\rho$ bounded and positive if $3\,k - 1 \in \mathbb{N}_0$.
In the case of asymptotic radiation the systems reads 
\[
(\dot{\Omega})^2 = \frac{\lambda}{3}
- \frac{R[\hat{h}]}{6}\,\Omega^2 + 
\Omega^{4}\,\frac{\rho}{3},
\quad \quad 
\dot{\rho} = - \Omega^{4\,k-1}\,\rho^{1 + k}\,w^*(\Omega^4\,\rho))\,\dot{\Omega}.
\]
It can be integrated across $\Omega = 0$ with $\rho$ bounded and positive if $4\,k - 1 \in \mathbb{N}_0$.

\vspace{.2cm}

In the following sections we derive the conformal 
Einstein-$\lambda$-perfect fluid equations, introduce a gauge that involves in particular an orthonormal frame $e_j$, $j = 0, 1, 2, 3$, with $e_0 = U$, and discuss the regularity  as $\Omega \rightarrow 0$ of the equations
in this gauge with any of the two asymptotic equations of state assumed.

In the case of an
asymptotic dust equations of state there arise problems.
The use of the asymptotic equation of state affects the principal part of the fluid equations in a way which does not allow us to apply 
the methods which were successful  in the case of the pure dust equation of state \cite{friedrich:dust:2016}. 
This case, which I consider as particularly interesting
(see \cite{friedrich:2023a}), is left open.

In the case of an asymptotic radiation equation of state 
regularity of the equations can be ensured by a suitable assumption 
and there can be derived a symmetric hyperbolic reduced system that is well defined irrespective of the sign of $\Omega$. This allows us to give the following answer to the question posed in the beginning.

\vspace{.2cm}

{\it Consider the reduced system of the conformal 
Einstein-$\lambda$-perfect-fluid equations in the gauge discussed below with an asymptotic radiation equation of state where
$k \ge 1$. On a compact 3-dimensional manifold 
${\cal J}^+$ let be given smooth Cauchy data for the reduced equations with 
$\Omega = 0$, $U^k = \delta^k\,_0$ time-like future directed, 
and $U^k\,\nabla_k \Omega < 0$ that satisfy the constraints induced by the conformal field equations and the special properties implied on a space-like hypersurface on which $\Omega = 0$. These data determine a smooth solution to the reduced equations with $\Omega < 0$ in the future of 
${\cal J}^+$ and $\Omega > 0$ in the past of ${\cal J}^+$. Where 
$\Omega > 0$ the solution defines a unique solution to the 
system of the Einstein-$\lambda$-perfect-fluid equations with an asymptotic radiation equation of state 
that is time-like geodesically future complete and for which 
${\cal J}^+$ represents the conformal boundary at the infinite time-like future. 

\vspace{.1cm}

Let $S$ be a Cauchy hypersurface for this solution in the past of 
${\cal J}^+$ and denote by $\Delta$ the Cauchy data induced by the solution on $S$. Any Cauchy data $\Delta'$ on $S$ for the same system which are sufficiently close to $\Delta$ develop into a solution which is
also time-like geodesically future complete, admits a smooth conformal boundary in its future, and a smooth conformal extension beyond. }

\vspace{.2cm}

At the end of this article are discussed observations about solutions for which $U$ is orthogonal to the hypersurface ${\cal J}^+$ and consequently hypersurface orthogonal on physical solution in the past of ${\cal J}^+$.

\section{The conformal Einstein-$\lambda$-Euler system.}

We consider the Einstein-$\lambda$-perfect fluid system with cosmological constant $\lambda$ that consists of the Einstein equations (\ref{einst}) with cosmological constant $\lambda > 0$ and an energy momentum tensor of a simple ideal fluid
given by
\[
\hat{T}_{\mu \nu} = (\hat{\rho} + \hat{p})\,\hat{U}_{\mu}\,\hat{U}_{\nu} + \hat{p}\,\hat{g}_{\mu \nu}.
\]
The unknowns are a Lorentz metric $\hat{g}_{\mu \nu}$ on a four-dimensional manifold $\hat{M}$, a future directed time-like fluid flow vector field $\hat{U}^{\mu}$ satisfying $\hat{U}_{\mu}\,\hat{U}^{\mu} = - 1$, the total energy density $\hat{\rho}$ and the pressure  $\hat{p}$ 
as measured by co-moving  observers. The system is be completed by an equation of state 
\begin{equation}
\label{eos}
\hat{p} = w(\hat{\rho}) \ge 0.
\end{equation}
The matter equations, equivalent to $\hat{\nabla}^{\mu}\,\hat{T}_{\mu \nu} = 0$, are given by
\begin{equation}
\label{energy}
\hat{U}^{\mu}\,\hat{\nabla}_{\mu}\,\hat{\rho} 
+ (\hat{\rho} + \hat{p})\,\hat{\nabla}_{\mu}\,\hat{U}^{\mu} = 0,
\end{equation}
\begin{equation}
\label{flow}
(\hat{\rho} + \hat{p})\,\hat{U}^{\mu}\,\hat{\nabla}_{\mu}\,\hat{U}_{\nu}
+ \{\hat{U}^{\mu}\,\hat{U}_{\nu} + \hat{g}^{\mu}\,_{\nu}\}\,\hat{\nabla}_{\mu}\hat{p}  = 0.
\end{equation}

\vspace{.1cm}

We assume that $\hat{\rho} > 0$ on an initial Cauchy slice. 
The first of the matter equations then ensures that this relation is preserved where the solution to the equations is regular. 
Let $\hat{e}_a$, $a = 1, 2, 3$, be vector fields that satisfy
$\hat{g}(\hat{U}, \hat{e}_a) = 0$,  $\hat{g}( \hat{e}_a, \hat{e}_b) = 0$
and $F = F(\hat{\rho})$ a function that satisfies
\begin{equation}
\label{def-F}
F' = - (\hat{\rho} + w)^{-1}\,w',
\end{equation}
where we use, as in the following chapters, the notation
$' = \frac{\partial}{\partial \hat{\rho}}$.
With (\ref{flow}) follows  
\begin{equation}
\label{hat-f-a}
\hat{f}_a \equiv \hat{U}^{\mu}\,
\hat{e}^{\nu}\,_a\,\hat{\nabla}_{\mu}\,\hat{U}_{\nu} = \hat{e}_a(F).
 \end{equation}
Equations (\ref{energy}), (\ref{flow}) imply
\begin{equation}
\label{nabla-rho} 
w'\,\hat{\nabla}_{\nu}\hat{\rho}  = 
(\hat{\rho} + w)\,
\left(w'\,\hat{U}_{\nu}\,
\hat{\nabla}_{\rho}\,\hat{U}^{\rho}
- \hat{U}^{\rho}\,\hat{\nabla}_{\rho}\,\hat{U}_{\nu}\right).
\end{equation}
Taking a derivative and observing  that $\hat{\rho} + w > 0$ we 
get the equation
\[ 
w''\,\hat{\nabla}_{\mu}\hat{\rho} \,\,\hat{\nabla}_{\nu}\hat{\rho} 
+ w'\,\,\hat{\nabla}_{\mu}\,\hat{\nabla}_{\nu}\hat{\rho}   
= 
w'\,\frac{1 + w'}{\hat{\rho} + w}\,\hat{\nabla}_{\mu}\hat{\rho}
\,\hat{\nabla}_{\nu}\hat{\rho}
\]
\[
+ (\hat{\rho} + w)\left(
w''\,\hat{\nabla}_{\mu}\hat{\rho}\,\,\hat{U}_{\nu}\,
\hat{\nabla}_{\rho}\,\hat{U}^{\rho}
+ w'\,\hat{\nabla}_{\mu}\,\hat{U}_{\nu}\,
\hat{\nabla}_{\rho}\,\hat{U}^{\rho}
+ w'\,\hat{U}_{\nu}\,\hat{\nabla}_{\mu}\,
\hat{\nabla}_{\rho}\,\hat{U}^{\rho}
\right)
\]
\[
- (\hat{\rho} + w)\left(
\hat{\nabla}_{\mu}\hat{U}^{\rho}\,\hat{\nabla}_{\rho}\,\hat{U}_{\nu}
 + \hat{U}^{\rho}\,\hat{\nabla}_{\mu}\hat{\nabla}_{\rho}\,\hat{U}_{\nu}
\right).
\]
Where $w' \neq 0$ anti-symmetrization of this relation implies the integrability condition
\begin{equation}
\label{2nd-U-rho-int-cond}
\hat{U}^{\rho}\,\hat{\nabla}_{\rho}\hat{\nabla}_{[\mu}\,\hat{U}_{\nu]} 
- \hat{\nabla}_{\rho}\,\hat{U}_{[\mu}\,\hat{\nabla}_{\nu]}\hat{U}^{\rho}
= 
\end{equation}
\[
w'\,\hat{\nabla}_{[\mu}\,\hat{U}_{\nu]}\,
\hat{\nabla}_{\rho}\,\hat{U}^{\rho}
- w'\,\hat{U}_{[\mu}\,\hat{\nabla}_{\nu]}\,
\hat{\nabla}_{\rho}\,\hat{U}^{\rho}
- w''\,\frac{\hat{\rho} + w}{w'}\,\hat{\nabla}_{\lambda}\,\hat{U}^{\lambda}\,
\hat{U}^{\rho}\,\hat{\nabla}_{\rho}\,\hat{U}_{[\mu}\,\hat{U}_{\nu]}.
\]

\vspace{.1cm}

\noindent
{\bf Conformal transformations of the curvature fields.}

\vspace{.2cm}

\noindent
The conformal Weyl tensor $C^{\mu}\,_{\nu \lambda \eta}$ and the Schouten tensor $L_{\mu\nu} = 1/2\left(R_{\mu\nu} - 1/6\,R\, 
g_{\mu\nu}\right)$ of a given metric satisfy the contracted Bianchi  identity
\[
\nabla_{\mu}\,C^{\mu}\,_{\nu \lambda \eta} 
= 2\,\nabla_{[\lambda}\,L_{\eta]\nu}.
\]
Under the rescaling
\[
\hat{g}_{\mu\nu} \rightarrow g_{\mu\nu} = \Omega^2\,\hat{g}_{\mu\nu}, \quad \Omega > 0,
\]
the conformal Weyl tensor is invariant,
$C^{\mu}\,_{\nu \lambda \eta} = \hat{C}^{\mu}\,_{\nu \lambda \eta}$, 
 the Ricci tensor transform as
\[
\hat{R}_{\mu\nu} \rightarrow
R_{\mu\nu} = \hat{R}_{\mu\nu}
- 2\,\Omega^{-1}\nabla_{\mu}\nabla_{\nu}\Omega
- g^{\lambda \delta} 
\left(
\Omega^{-1}\,\nabla_{\lambda}\nabla_{\delta}\Omega
- 3\,\Omega^{-2}\nabla_{\lambda}\Omega\nabla_{\delta}\Omega
\right)g_{\mu\nu}, 
\]
and the Schouten tensor as
\begin{equation}
\label{schouten-conf-traf}
\hat{L}_{\mu\nu}  \rightarrow
L_{\mu\nu} = \hat{L}_{\mu\nu} 
- \Omega^{-1}\nabla_{\mu}\nabla_{\nu}\Omega
+ 1/2\,\,\Omega^{-2}\,g^{\lambda\delta}\,
\nabla_{\lambda}\Omega\,\nabla_{\delta}\Omega\,g_{\mu\nu}. 
\end{equation}
The conformal Weyl tensor satisfies
$\nabla_{\mu}\left(\Omega^{-1}\, C^{\mu}\,_{\nu \lambda \eta}\right)
= \Omega^{-1}\,\hat{\nabla}_{\mu}\,\hat{C}^{\mu}\,_{\nu \lambda \eta}$
where $\nabla$ and $\hat{\nabla}$ denote the Levi-Civita connections of $g_{\mu\nu}$ and  $\hat{g}_{\mu\nu}$.
The identity above thus implies for the 
{\it rescaled conformal Weyl tensor} 
$W^{\mu}\,_{\nu \lambda \eta} = \Omega^{-1}\, C^{\mu}\,_{\nu \lambda \eta}$ the relation
\begin{equation}
\label{resc-Weyl-equ}
\nabla_{\mu}\,W^{\mu}\,_{\nu \lambda \eta} 
= 2\,\Omega^{-1}\,\hat{\nabla}_{[\lambda}\,\hat{L}_{\eta]\nu},
\end{equation}
whence
\begin{equation}
\label{conf-Shouten-equ}
2\,\nabla_{[\lambda}\,L_{\eta]\nu} - \nabla_{\mu}\,\Omega\,W^{\mu}\,_{\nu \lambda \eta} 
= 2\,\hat{\nabla}_{[\lambda}\,\hat{L}_{\eta]\nu}.
\end{equation}

\vspace{.2cm}

\noindent
{\bf Conformal transformation of the matter fields.}

\vspace{.2cm}

\noindent
We combine the conformal rescaling of the metric with the rescalings 
\[
U^{\mu} = \Omega^{-1}\,\hat{U}^{\mu} \quad
U_{\mu} = \Omega\,\hat{U}_{\mu}, 
\quad \rho = \Omega^{- e}\,\hat{\rho},
\quad \mbox{with $e = const. > 0$.}
\]
In the following it will always be understood that
\[
w = w(\hat{\rho}) = w(\Omega^{e}\,\rho), \quad \quad 
w' = w'(\hat{\rho}) = w'(\Omega^{e}\,\rho),
\]
and similarly with the function $w^*$ introduced later.
Equations (\ref{einst}) imply
\[
\hat{T} = \hat{g}^{\mu\nu}\,\hat{T}_{\mu\nu}
= 3\,w - \hat{\rho}, \quad \quad
\hat{R} = 4\,\lambda - \hat{T} = 
4\,\lambda + \hat{\rho} - 3\,w,
\]
and  
$\hat{L}_{\mu\nu} = \hat{L}^*_{\mu\nu}
+ \frac{1}{6}\,\lambda\,\hat{g}_{\mu\nu}$
with
\begin{equation}
\label{L*}
\hat{L}^*_{\mu\nu} =
\frac{1}{2}\,(\hat{\rho} + w)\,\hat{U}_{\mu}\,\hat{U}_{\nu}  
+ \frac{1}{6}\,\hat{\rho}\,\hat{g}_{\mu\nu}
= \Omega^{- 2}\left(
\frac{1}{2}\,(\hat{\rho} +  w)\,U_{\mu}\,U_{\nu}  
+ \frac{1}{6}\,\hat{\rho}\,g_{\mu\nu}
\right),
\end{equation}
It follows that
$\hat{\nabla}_{\lambda}\hat{L}_{\eta\nu} = 
\hat{\nabla}_{\lambda}\hat{L}^*_{\eta\nu}$. 
The Ricci scalar of the conformal metric satisfies
\[
6\,\Omega\,\nabla_{\mu}\nabla^{\mu}\Omega + \Omega^2\,R 
- 12\,\nabla_{\mu}\Omega\nabla^{\mu}\Omega
= 4\,\lambda + \hat{\rho} - 3\,w.
\]
Written in the form
\begin{equation}
\label{conf-lambda-equ}
6\,\Omega\,s 
- 3\,\nabla_{\mu}\Omega\nabla^{\mu}\Omega
= \lambda + \frac{1}{4}\,(\hat{\rho} - 3\,w)
 \quad \mbox{with} \quad 
s =  \frac{1}{4}\,\nabla_{\mu}\nabla^{\mu}\Omega
+ \frac{1}{24}\,\Omega\,R,
\end{equation}
it will be referred to as {\it $\lambda$-equation}.
Equations (\ref{schouten-conf-traf}), (\ref{L*}),
(\ref{conf-lambda-equ}) give
\begin{equation}
\label{nabla-nabla-Omega-equ}
\nabla_{\mu}\nabla_{\nu}\Omega
= - \Omega\,L_{\mu\nu} + s\,g_{\mu\nu}
+ \Omega^{-1}\,\frac{\hat{\rho} + w}{2}\left(
U_{\mu}\,U_{\nu} + \frac{1}{4}\,g_{\mu\nu}\right)
\end{equation}
which will be referred to as {\it $\Omega$-equation}.
Applying a derivative to the $\lambda$-equation (\ref{conf-lambda-equ})
gives with (\ref{nabla-nabla-Omega-equ}) the {\it $s$-equation}
\begin{equation}
\label{nabla-s}
\nabla_{\mu} s + L_{\mu\nu}\,\nabla^{\nu}\Omega =
\Omega^{-2}\,\frac{\hat{\rho} + w}{2}\left(
U_{\mu}\,U^{\nu}\,\nabla_{\nu}\Omega 
+ \frac{1}{4}\,\nabla_{\mu}\Omega\right)
+ \frac{1}{24}\,(1 - 3\,w')\Omega^{-1}\,\nabla_{\mu}\hat{\rho}.
\end{equation}
It follows
\[
\hat{\nabla}_{[\lambda}\,\hat{L}_{\eta]\nu} =
\nabla_{[\lambda}\,\hat{L}^*_{\eta]\nu}
- \Omega^{-1}\left(
\hat{L}^*_{\nu[\lambda}\,\nabla_{\eta]}\,\Omega
+ g_{\nu [\lambda }\, \hat{L}^*_{\eta] \rho}\,\nabla^{\rho}\Omega
\right),
\]
which gives with (\ref{L*}) and  $\nabla_{\lambda}\hat{\rho} =
\Omega^{e}\,(\nabla_{\lambda}\rho 
+ e\,\rho\,\Omega^{-1}\,\nabla_{\lambda}\Omega)$ 
the relation
\begin{equation}
\label{nabla-hat-L-hat}
2\,\hat{\nabla}_{[\lambda}\,\hat{L}_{\eta]\nu} = 
\Omega^{-3}\,(\hat{\rho} + w)\left(
 U_{\nu}\,U_{[\lambda}\,\nabla_{\eta]}\,\Omega
 - g_{\nu [\lambda }\,U_{\eta]}\,U_{ \rho}\,\nabla^{\rho}\Omega
\right)
\end{equation}
\[
+ e\,\rho\,\,\Omega^{e - 3}\left( (1 + w')\,
\nabla_{[\lambda}\Omega\,U_{\eta]}\,U_{\nu}
+ \frac{1}{3}\,\nabla_{[\lambda}\Omega\,g_{\eta]\nu}\right)
\]
\[
+ \Omega^{-2}\,
(\hat{\rho} + w)
\left(\nabla_{[\lambda}\,U_{\eta]}\,U_{\nu}  
+ U_{[\eta}\,\nabla_{\lambda]}\,U_{\nu}  
\right)
\]
\[
+ \Omega^{e-2}\,\left((1 + w')\,
\nabla_{[\lambda}\,\rho\,\,U_{\eta]}\,U_{\nu}  
+ \frac{1}{3}\,\nabla_{[\lambda}\,\rho\,g_{\eta]\nu}
\right).
\]

\vspace{.2cm}

\noindent
{\bf The conformal matter equations}

\vspace{.2cm}

\noindent
Equations (\ref{energy}) and (\ref{flow}) transform with 
$\nabla_{\mu}\hat{\rho} = 
\Omega^{e}\left\{\nabla_{\mu}\,\rho 
+  e\,\rho\,\Omega^{-1}\,\nabla_{\mu}\Omega \right\}$
into 
\begin{equation}
\label{conf-w-energy}
0 = U^{\mu}\,\nabla_{\mu}\rho
+ (\rho + \Omega^{-e}\,w)\,\nabla_{\mu}\,U^{\mu}
- \Omega^{-1}\,(- e\,\rho + 3\,\rho + 3\,\Omega^{-e}\,w)\,U^{\mu}\,\nabla_{\mu}\Omega,
\end{equation}
\begin{equation}
\label{conf-w-flow}
0 = (\rho + \Omega^{-e}\,w)\,U^{\mu}\,\nabla_{\mu}\,U_{\nu}
+ w'\,
(g^{\mu}\,_{\nu} + U^{\mu}\,U_{\nu})\,\nabla_{\mu}\,\rho
\quad \quad \quad 
\quad \quad \quad 
\quad \,\,\,
\end{equation}
\[
\quad \quad \quad 
\quad \quad 
- \Omega^{-1}\,(\rho + \Omega^{-e}\,w
- e\,\rho\,w')
(g^{\mu}\,_{\nu} + U^{\mu}\,U_{\nu})\nabla_{\mu}\Omega.
\]

\vspace{.2cm}

\noindent
In the case of pure dust, where 
$w = 0$ and $e = 3$, the equations read
\[
0 = U^{\mu}\,\nabla_{\mu}\rho
+ \rho\,\nabla_{\mu}\,U^{\mu},
\quad \quad
0 = \rho\,U^{\mu}\,\nabla_{\mu}\,U_{\nu}
- \Omega^{-1}\, \rho\,
(g^{\mu}\,_{\nu} + U^{\mu}\,U_{\nu})\nabla_{\mu}\Omega.
\]
The $\Omega^{-1}$ term in the second equation reflects the conformal non-covariance of the system.  
In the case of pure radiation with $w = \frac{1}{3}\,\hat{\rho}$,
and $e = 4$ the conformal equations reduce to
\[
U^{\mu}\,\nabla_{\mu}\rho
+ \frac{4}{3}\,\rho\,\nabla_{\mu}\,U^{\mu} = 0,
\quad \quad
\frac{4}{3}\,\rho\,\,U^{\mu}\,\nabla_{\mu}\,U_{\nu}
+ \frac{1}{3}\,
(g^{\mu}\,_{\nu} + U^{\mu}\,U_{\nu})\,\nabla_{\mu}\,\rho
= 0,
\]
and have thus the same form as their physical versions. 

\vspace{.1cm}

\noindent
Equation 
(\ref{2nd-U-rho-int-cond}) transforms into
\begin{equation}
\label{conf-2nd-U-rho-int-cond} 
U^{\rho}\,\nabla_{\rho}\nabla_{[\mu}\,U_{\nu]} 
+ w'\,U_{[\mu}\,\nabla_{\nu]}\,\nabla_{\rho}\,U^{\rho} 
\end{equation}
\[
- \nabla_{\pi}\,U_{[\mu}\,\nabla_{\nu]}\,U^{\pi}  
- (1 - 3\,w')\,U_{[\mu}\,L_{\nu]\rho}\,U^{\rho}
- w'\,\nabla_{[\mu}\,U_{\nu]}\,\nabla_{\pi}\,U^{\pi} =
\]
\[
+ (1 - 3\,w')\left\{\Omega^{-1}\,\nabla_{[\mu}\,U_{\nu]}\,U^{\rho}\,\nabla_{\rho}\Omega 
- \Omega^{-1}\,U_{[\mu}\,\nabla_{\nu]}\,U_{\rho}\,\nabla^{\rho}\Omega
+ \Omega^{-2}\,U_{[\mu}\,\nabla_{\nu]}\Omega\,
U^{\rho}\,\nabla_{\rho}\Omega \right\} 
\]
\[
- w''\,\frac{\hat{\rho} + w}{w'}\,
\left\{\nabla_{\pi}\,U^{\pi}
\,U^{\rho}\,\nabla_{\rho}\,U_{[\mu}\,U_{\nu]}  
- 3\,\Omega^{-1}\,U^{\pi}\,\nabla_{\pi}\Omega\,U^{\rho}\,\nabla_{\rho}\,U_{[\mu}\,U_{\nu]}\right. 
\]
\[
\left.- \Omega^{-1}\,\nabla_{\pi}\,U^{\pi}\,\nabla_{[\mu}\Omega\,U_{\nu]}
+ 3\,\Omega^{-2}\,U^{\pi}\,\nabla_{\pi}\Omega\,\nabla_{[\mu}\Omega\,U_{\nu]}\right\}.
\]

\vspace{.1cm}

\noindent
With $\hat{e}^{\mu}\,_a = \Omega\,e^{\mu}\,_a$
the relation (\ref{hat-f-a}) transforms into 
\begin{equation}
\label{hat-f-a--f-a}
e_a(F) = \Omega^{-1}\,\hat{f}_a 
= U^{\mu}\,\nabla_{\mu}U_{\nu}\,e^{\nu}\,_a 
- e_a(\log \Omega).
 \end{equation}

\section{The gauge and the implied equations.}

We express the equations in terms of an orthonormal frame field
$e_k =e^{\mu}\,_k\partial_{x^{\mu}}$, $k = 0, 1, 2, 3$, so that $g_{jk} \equiv g(e_j, e_k) = \eta_{jk} = diag(-1, 1, 1, 1)$ and $e_0$
is a time-like vector field. 
The space-like frame vector fields are then given by the $e_a$, where $a, b, c = 1, 2, 3$ denote spatial indices to which the summation convention also applies. 
In the following all tensor fields considered in the previous sections will be expressed in terms of this frame field.
The contravariant coordinate version of the metric is given by 
$g^{\mu \nu} = \eta^{jk}\,e^{\mu}\,_j\,e^{\nu}\,_k$. 

The connection coefficients, defined by 
$\nabla_je_k \equiv \nabla_{e_j}e_k = \Gamma_j\,^l\,_k\,e_l$, 
satisfy $ \Gamma_{jlk} = - \Gamma_{jkl}$ with $ \Gamma_{jlk} = \Gamma_j\,^i\,_k\,g_{li}$,
because $\nabla_i g_{jk} = 0$. 
The covariant derivative of a tensor field $X^{\mu}\,_{\nu}$, given in the frame by 
$X^i\,_j$, takes the form
$\nabla_k\,X^i\,_j =  X^i\,_{j\,,\mu}\,e^{\mu}\,_k +  \Gamma_k\,^i\,_l \,X^l\,_j - 
\Gamma_k\,^i\,_l\,X^i\,_j$.
The frame and the connection coefficients satisfy the {\it first structural equations} 
\begin{equation}
\label{conf-O-torsion-free condition}
e^{\mu}\,_{i,\,\nu}\,e^{\nu}\,_{j}
 - e^{\mu}\,_{j,\,\nu}\,e^{\nu}\,_{i} 
= (\Gamma_{j}\,^{k}\,_{i} - \Gamma_{i}\,^{k}\,_{j})\,e^{\mu}\,_{k},
\end{equation}
which ensures that the connection is torsion free, and  the {\it second structural equations}
\begin{equation}
\label{conf-O-Ricci identity}
\Gamma_l\,^i\,_{j,\,\mu}\,e^{\mu}\,_k - \Gamma_k\,^i\,_{j,\,\mu}\,e^{\mu}\,_l
+ 2\,\Gamma_{[k}\,^{i\,p}\,\Gamma_{l]pj}
- 2\,\Gamma_{[k}\,^p\,_{l]}\,\Gamma_p\,^i\,_j
\end{equation}
\[
= \Omega\,W^i\,_{jkl}
+ 2\,\{g^i\,_{[k}\,L_{l] j} + L^i\,_{ [k}\,g_{l] j}\}.
\]
To restrict the gauge freedom for the frame we set 
$e_0 = U$
so that 
$U = U^k\,e_k$
with $U^k = \delta^k\,_0$,
choose at the points of a given smooth space-like hypersurface $S$ transverse to the flow line of $U$ vector fields $e_a$, $a = 1, 2, 3$, so that
$g(e_j, e_k) = \eta_{jk}$, assume the orthonormal frame $e_k$ 
to be extended by Fermi-transport in the direction of $e_0 = U$ so that
$0 = \mathbb{F}_U e_k = \nabla_U e_k - g(e_k,  \nabla_U U)\,U 
+ g(e_k, U)\, \nabla_U U$, and assume the $e_a$ to be chosen on $S$ so that the resulting orthonormal frame field is smooth.
In terms of this frame the Fermi transport law reduces to
\begin{equation}
\label{conf-Fermi-cond}
\Gamma_0\,^a\,_b = 0.
\end{equation}
To restrict the gauge freedom  for the coordinates $x^{\mu}$, we assume 
that $\tau \equiv x^0 = 0$ on $S$ and the $x^{\mu}$ be
dragged along with $U$ so that
\[
<e_0, dx^{\mu}>\,= e^{\mu}\,_0 = U^{\mu} = \delta^{\mu}\,_0.
\]
The remaining non-vanishing connection coefficients are given by
\[
f_a \equiv \Gamma_0\,^0\,_a = \delta_{ab}\,\,\Gamma_0\,^b\,_0, \quad \quad
\chi_{ab} \equiv \Gamma_a\,^0\,_b = \delta_{bc}\,\,\Gamma_a\,^c\,_0 \quad \mbox{and} \quad 
\Gamma_a\,^b\,_c.
\]
It holds 
$\nabla_i\,U_k = \Gamma_i\,^0\,_k  = 
\chi_{ab}\,\delta^a\,_i\,\delta^b\,_k
+ f_{b}\,\delta^0\,_i\,\delta^b\,_k$ and
 $U^i\,\nabla_i\,U_k = f_{b}\,\delta^b\,_k$, 
$\nabla_i\,U^i = \chi_{a}\,^{a}$.

\noindent
The first structural equations supply the constraint
\begin{equation}
\label{s-frame-coeff-constr}
e^{\mu}\,_{a,\,\nu}\,e^{\nu}\,_{b}
 - e^{\mu}\,_{b,\,\nu}\,e^{\nu}\,_{a} 
= (\Gamma_{b}\,^{c}\,_{a} - \Gamma_{a}\,^{c}\,_{b})\,e^{\mu}\,_{c}
+ (\chi_{b a} - \chi_{a b})\,\delta^{\mu}\,_{0},
\end{equation}
and the evolution equations 
\begin{equation}
\label{s-frame-coeff-evol}
e^{\mu}\,_{a,\,0}
= - \chi_{a}\,^{b}\,e^{\mu}\,_{b}
+  f_{a}\,\delta^{\mu}\,_{0}.
\end{equation}
The second structural equations
imply for $\Gamma_a\,^b\,_{c}$ and
 $\chi_{ab}$ the constraints
 \begin{equation}
\label{s0a-Gamma-constr}
\Gamma_b\,^a\,_{c,\,\mu}\,e^{\mu}\,_d - \Gamma_d\,^a\,_{c,\,\mu}\,e^{\mu}\,_b
+ 2\,\Gamma_{[d}\,^{a\,i}\,\Gamma_{b]ic}
- 2\,\Gamma_{[d}\,^i\,_{b]}\,\Gamma_i\,^a\,_c
\end{equation}
\[
= \Omega\,W^a\,_{cdb}
+ 2\,\{g^a\,_{[d}\,L_{b] c} + L^a\,_{ [d}\,g_{b] c}\},
\]
\begin{equation}
\label{s0b-Gamma-constr}
\chi_{a b,\,\mu}\,e^{\mu}\,_c 
- \chi_{c b,\,\mu}\,e^{\mu}\,_a
+ 2\,\Gamma_{[c}\,^{0\,p}\,\Gamma_{a]pb}
- 2\,\Gamma_{[c}\,^p\,_{a]}\,\Gamma_p\,^0\,_b
= \Omega\,W^0\,_{bca} + 2\,L^0\,_{ [c}\,g_{a] b},
\end{equation}
and the evolution equations 
\begin{equation}
\label{s1-Gamma-evol}
\Gamma_a\,^b\,_{c,\,0} 
+ f^b\,\chi_{ac} - \chi_a\,^b\,f_c
+ \chi_{a}\,^d\,\Gamma_d\,^b\,_c
= \Omega\,W^b\,_{c0a}
- g^b\,_{a}\,L_{0 c} + L^b\,_{ 0}\,g_{a c},
\end{equation}
\begin{equation}
\label{s2b-Gamma-evol}
\chi_{ac,\,0} 
- f_{c,\,\mu}\,e^{\mu}\,_a + f_b\,\Gamma_a\,^b\,_c
- f_a\,f_b + \chi_a\,^b\,\chi_{bc}
= \Omega\,W^0\,_{c0a}
+ L_{a c} + L^0\,_{ 0}\,g_{a c}.
\end{equation}
No equation for $f_a$ is implied by the structural equations. 

\vspace{.1cm}

\noindent
The fluid equations 
(\ref{conf-w-energy}), (\ref{conf-w-flow}) imply in our gauge the evolution equation
\begin{equation}
\label{E-conf-w-energy}
\nabla_{0}\rho
+ (\rho + \Omega^{-e}\,w)\,\chi_{a}\,^{a}
- \Omega^{-1}\,(- e\,\rho + 3\,\rho + 3\,\Omega^{-e}\,w)\,\nabla_{0}\Omega = 0,
\end{equation}
and the constraint
\begin{equation}
\label{E-conf-w-flow}
(\rho + \Omega^{-e}\,w)\,f_a
+ w'\,\nabla_{a}\,\rho
- \Omega^{-1}\,(\rho + \Omega^{-e}\,w
- e\,\rho\,w')
\nabla_{a}\Omega = 0.
\end{equation}
With $\hat{\rho} = \Omega^{e}\,\rho $ the latter is seen to be (\ref{hat-f-a--f-a}), which reads with 
$f_a = U^{\mu}\,\nabla_{\mu}U_{\nu}\,e^{\nu}\,_a $
\begin{equation}
\label{f_a-meaning}
f_a = e_a(F + \log \Omega).
\end{equation}
Equation (\ref{conf-2nd-U-rho-int-cond}) 
is equivalent to the two equations
\begin{equation}
\label{0a-conf-2nd-U-rho-int-cond}
e_0(f_a)  - w'\,e_a(\chi_c\,^c) + \chi_{ac}\,f^c + (1 - 3\,w')\,L_{a0} 
- w'\,\chi_c\,^c\,f_a =
\end{equation}
\[
= (1 - 3\,w')\,(\Omega^{-1}\,\nabla_0\Omega\,f_a 
+ \Omega^{-1}\,\chi_{ab}\,\nabla^b\Omega  
- \Omega^{-2}\,\nabla_0\Omega\,\nabla_a\Omega)
\]
\[
- w''\,\frac{\hat{\rho} + w}{w'}\,(\chi_c\,^c\,f_a 
- 3\,\Omega^{-1}\,\nabla_0\,\Omega\,f_a 
- \Omega^{-1}\,\chi_c\,^c\,\nabla_a\Omega
+ \Omega^{-2}\,\nabla_0\Omega\,\nabla_a\Omega),
\]
and
\begin{equation}
\label{ab-conf-2nd-U-rho-int-cond}
e_0(\chi_{[ab]}) - \chi_{c[a}\,\chi_{b]}\,^c - w'\,\chi_c\,^c\,\chi_{[ab]}
= (1 - 3\,w')\,\Omega^{-1}\,\nabla_0\Omega\,\chi_{[ab]}.
\end{equation}
With $\chi_{c[a}\,\chi_{b]}\,^c = - \chi_{[ac]}\,\sigma_b\,^c
+  \chi_{[bc]}\,\sigma_a\,^c$, where $\sigma_{ab} =\chi_{(ab)}$, the latter can be read as a linear homogeneous ODE for $\chi_{[ab]}$.

\vspace{.1cm}

\noindent
The system of conformal field equations reads 
with $e = const.$
\begin{equation}
\label{f-G-coord-alg-equ}
6\,\Omega\,s - 3\,\nabla_{k}\Omega\,\nabla^{k}\Omega - 
\lambda = \frac{1}{4}\,(\hat{\rho} - 3\,w),
\end{equation}

\begin{equation}
\label{f-G-coord-Omega-equ}
\nabla_{k}\,\nabla_{j}\Omega + \,\Omega\,L_{kj} - s\,g_{kj}
= \Omega^{-1}\,\frac{\hat{\rho} + w}{2}\left(
U_{k}\,U_{j} + \frac{1}{4}\,g_{kj}\right),
\end{equation}

\vspace{.1cm}

\begin{equation}
\label{f-G-coord-s-equ}
\nabla_{k}\,s + \nabla^{i}\Omega\,L_{i k} = 
\end{equation}
\[
\Omega^{-2}\,\frac{\hat{\rho} + w}{2}\left(
U_{k}\,U^{i}\,\nabla_{i}\Omega 
+ \frac{1}{4}\,\nabla_{k}\Omega\right)
+ \frac{1}{24}\,(1 - 3\,w')\,
\Omega^{e- 1}\left\{\nabla_{\mu}\,\rho 
+ e\,\rho\,\Omega^{-1}\,\nabla_{k}\Omega \right\},
\]

\vspace{.1cm} 

\begin{equation}
\label{f-conf-G-coord-L-equ}
\nabla_{i}\,L_{j k} 
- \nabla_{j}\,L_{i k} - 
\nabla_{l}\Omega\,\,W^{l}\,_{k i j} 
= 2\,\hat{\nabla}_{[i}\,\hat{L}_{j] k},
\end{equation}

\vspace{.1cm}

\begin{equation}
\label{f-conf-G-coord-W-equ}
\nabla_{l}\,W^{l}\,_{k i j} 
= 2\,\Omega^{-1}\,\hat{\nabla}_{[i}\,\hat{L}_{j] k},
\end{equation}

\vspace{.1cm}

\noindent
with
\begin{equation}
\label{ON-nabla-hat-L-hat}
2\,\hat{\nabla}_{[i}\,\hat{L}_{j]k} = 
\Omega^{-3}\,(\hat{\rho} + w)\left(
 U_{k}\,U_{[i}\,\nabla_{j]}\,\Omega
 - g_{k [i }\,U_{j]}\,U^{l}\,\nabla_{l}\Omega
\right)
\end{equation}
\[
+ e\,\rho\,\Omega^{e - 3}\left( (1 + w')\,
\nabla_{[i}\Omega\,U_{j]}\,U_{k}
+ \frac{1}{3}\,\nabla_{[i}\Omega\,g_{j]k}\right)
\]
\[
+ \Omega^{-2}\,
(\hat{\rho} + w)
\left(\nabla_{[i}\,U_{j]}\,U_{k}  
+ U_{[j}\,\nabla_{i]}\,U_{k}  
\right)
\]
\[
+ \Omega^{e-2}\,\left((1 + w')\,
\nabla_{[i}\,\rho\,\,U_{j]}\,U_{k}  
+ \frac{1}{3}\,\nabla_{[i}\,\rho\,g_{j]k}
\right).
\]

\section{Regularity of the equations.}

In the following an equation will be called `regular' if no negative or non-integer powers of $\Omega$ occur in it. The structural equations are regular in this sense but the tensorial equations can be problematic.
It will use from now on the notation
\begin{equation}
\label{def-Sigma}
\nabla_k\Omega = \Sigma_k.
\end{equation} 

\vspace{.1cm}

\noindent
{\bf The  asymptotic dust equation of state.}

\vspace{.1cm}

With the asymptotic dust equation of state and  $e = 3$ there enter in various places of the equations functions such as

\[
w = (\Omega^3\,\rho)^{k + 1}\,w^*, 
\quad \quad  
w' = \Omega^{3\,k}\, \rho^k\left( (k + 1)\,w^* 
+ \Omega^3 \rho\,\,(w^*)'\right),
\]
\[
\hat{\rho} + w = \Omega^{3}\,\rho\,(1 +
\Omega^{3\,k} \rho^{k}\,\,w^*), \quad \quad 
1 - 3\,w' = 1 - 3\,\Omega^{3\,k}\,\rho^k\left( (k + 1)\,w^* 
+ \Omega^3\,\rho\,(w^*)'\right).
\]
As $w^* \rightarrow 0$ they approach the values of the corresponding quantities of the pure dust equation of state. Therefore it is clear that equations which are singular in that case must also be singular in the present case.

 In the present case the singular terms cannot be handled as in the pure dust case because equation (\ref{flow}) 
remains
a partial differential equation and the $\Omega^{-1}$ terms cannot be compensated by suitable choices of $k$.
 Moreover, the expression
\[
w''\,\frac{\hat{\rho} + w}{w'} = (1 + \hat{\rho}^k\,w^*)
\left(k + \hat{\rho}\,
\frac{(2 + k)\,(w^*)' 
+ \hat{\rho}\,(w^*)''}{
(1 + k)\,w^* + \hat{\rho}\,(w^*)'}
\right),
\]
which comes with a singular factor in the equation
\[
e_0(f_a)  - w'\,e_a(\chi_c\,^c) + \chi_{ac}\,f^c + (1 - 3\,w')\,L_{a0} 
- w'\,\chi_c\,^c\,f_a =
\]
\[
= (1 - 3\,w')\,(\Omega^{-1}\,\Sigma_0\,f_a 
+ \Omega^{-1}\,\chi_{ab}\,\Sigma^b  
- \Omega^{-2}\,\Sigma_0\,\Sigma_a)
\]
\[
- w''\,\frac{\hat{\rho} + w}{w'}\,(\chi_c\,^c\,f_a 
- 3\,\Omega^{-1}\,\Sigma_0\,f_a 
- \Omega^{-1}\,\chi_c\,^c\,\Sigma_a
+ \Omega^{-2}\,\Sigma_0\,\Sigma_a),
\]
is not even defined if
$w' = 0$. Since there is no obvious way to handle it,  this case will not be considered any further in this article.

\vspace{.3cm}

\noindent
{\bf The  asymptotic radiation equation of state.}

\vspace{.3cm}

\noindent
The asymptotic radiation equation of state with  $e = 4$ gives rise to expressions like
\[
\hat{\rho} + w = \Omega^4\,\rho\,\left(\frac{4}{3} - (\Omega^4\,\rho)^{k}\,w^*\right), \quad \quad 
w' = \frac{1}{3} - (\Omega^{4}\, \rho)^k\left( (k + 1)\,w^* 
+ \Omega^4\, \rho\,(w^*)'\right),
 \]
 \[
\hat{\rho} - 3\,w = 
 3\,\Omega^{4\,(k + 1)}\,\rho^{k + 1}\,w^*(\Omega^{4}\,\rho),
 \quad \quad 
1 - 3\,w' = 3\,(\Omega^{4}\, \rho)^k\left( (k + 1)\,w^* 
+ \Omega^4\, \rho\,(w^*)'\right),
\]
\[
w''\,\frac{\hat{\rho} + w}{w'} = - (\Omega^4\,\rho)^k \left(4 - 3\,\hat{\rho}^k\,w^*\right)
\frac{k\,(1 + k)\,w^* + 2\,(1 + k)\,\hat{\rho}\,(w^*)' 
+ \hat{\rho}^2\,(w^*)''}{
1 - 3\,(1 + k)\,\hat{\rho}^k\,w^* + 3\,\hat{\rho}^{1 + k}\,(w^*)'}.
\]
The limits as $w^* \rightarrow 0$ are well defined and yield the corresponding functions in the pure radiation case. 
If $\rho > 0$ and bounded as $\Omega \rightarrow 0$ we can assume for small $|\Omega|$ that
\begin{equation}
\label{w'-pos-domain}
w' = \frac{1}{3} - \Omega^{4\,k}\, \rho^k\left( (k + 1)\,w^* 
+ \Omega^4\, \rho\,(w^*)'\right) > 0.
\end{equation}

\vspace{.2cm}

\noindent
The structural equation are regular with no condition on $k$. 
Equations (\ref{f-G-coord-alg-equ}), (\ref{f-G-coord-Omega-equ}), (\ref{f-G-coord-s-equ}) are immediately seen to be regular with
$4\,k \in \mathbb{N}_0$.
The fluid equation (\ref{E-conf-w-energy}) reads
 \begin{equation}
 \label{rad-E-conf-w-energy}
\nabla_{0}\rho
+ \rho\left\{\left(\frac{4}{3} -  (\Omega^4\,\rho)^{k}\,w^*\right)\,\chi_{a}\,^{a}
+ 3\,\Omega^{4\,k - 1}\rho^{k}\,w^*)\,\Sigma_{0}\right\} = 0.
\end{equation}
It is  regular if $4\,k \in \mathbb{N}$.
As long as the term in curly brackets is continuous, the function $\rho$ will stay positive if it  positive on some Cauchy surface. 
With (\ref{w'-pos-domain}) the constraint
(\ref{E-conf-w-flow}) can be solved for $\nabla_{a}\,\rho$ in a neighbourhood of $\Omega = 0$ 
\begin{equation}
\label{rad-E-conf-w-flow}
\nabla_{a}\,\rho 
= - \rho \,\frac{4 -  3\,(\Omega^4\,\rho)^{k}\,w^*}{3\,w'}\,f_a
- \Omega^{4\,k - 1}\,\rho^{k + 1}\,\frac{
(4\,k + 3)\,w^* + 4\,\Omega^{4}\,\rho\,(w^*)'}{w'}\,\Sigma_{a}.
\end{equation}
The equation is regular if $4\,k \in \mathbb{N}$.
Inspection of the equations
\begin{equation}
\label{0a-conf-2nd-U-rho-int-cond}
e_0(f_a)  - w'\,e_a(\chi_c\,^c) + \chi_{ac}\,f^c + (1 - 3\,w')\,L_{a0} 
- w'\,\chi_c\,^c\,f_a =
\end{equation}
\[
= (1 - 3\,w')\,(\Omega^{-1}\,\Sigma_0\,f_a 
+ \Omega^{-1}\,\chi_{ab}\,\Sigma^b  
- \Omega^{-2}\,\Sigma_0\,\Sigma_a)
\]
\[
- w''\,\frac{\hat{\rho} + w}{w'}\,(\chi_c\,^c\,f_a 
- 3\,\Omega^{-1}\,\Sigma_0\,f_a 
- \Omega^{-1}\,\chi_c\,^c\,\Sigma_a
+ \Omega^{-2}\,\Sigma_0\,\Sigma_a),
\]
and
\begin{equation}
\label{ab-conf-2nd-U-rho-int-cond}
e_0(\chi_{[ab]}) - \chi_{c[a}\,\chi_{b]}\,^c - w'\,\chi_c\,^c\,\chi_{[ab]}
= (1 - 3\,w')\,\Omega^{-1}\,\Sigma_0\,\chi_{[ab]},
\end{equation}
shows that they are regular if $4\,k - 1 \in \mathbb{N}$.
With the expressions above the equations
\begin{equation}
\label{rad-f-G-coord-alg-equ}
6\,\Omega\,s - 3\,\Sigma_{k}\,\Sigma^{k} - 
\lambda = \frac{1}{4}\,(\hat{\rho} - 3\,w),
\end{equation}
\begin{equation}
\label{rad-f-G-coord-Omega-equ}
\nabla_{k}\,\Sigma_{j} =  - \Omega\,L_{kj} + s\,g_{kj}
+ \Omega^{-1}\,\frac{\hat{\rho} + w}{2}\left(
U_{k}\,U_{j} + \frac{1}{4}\,g_{kj}\right),
\end{equation}
\begin{equation}
\label{rad-f-G-coord-s-equ}
\nabla_{k}\,s + \Sigma^{i}\,L_{i k} = 
\end{equation}
\[
\Omega^{-2}\,\frac{\hat{\rho} + w}{2}\left(
U_{k}\,U^{i}\,\Sigma_{i} 
+ \frac{1}{4}\,\Sigma_{k}\right)
+ \frac{1}{24}\,(1 - 3\,w')\,
\Omega^{3}\left\{\nabla_{k}\,\rho 
+ 4\,\rho\,\Omega^{-1}\,\Sigma_{k} \right\},
\]
are seen to be regular if $4\,k \in \mathbb{N}_0$.
We have finally
\begin{equation}
\label{rad-f-conf-G-coord-L-equ}
\nabla_{i}\,L_{j k} 
- \nabla_{j}\,L_{i k} - 
\Sigma_{l}\,W^{l}\,_{k i j} 
=  \Omega\,M_{ijk},
\end{equation}
\begin{equation}
\label{rad-f-conf-G-coord-W-equ}
\nabla_{l}\,W^{l}\,_{k i j} =  M_{ijk},
\end{equation}
where
\begin{equation}
\label{ON-nabla-hat-L-hat}
M_{ijk} = 2\,\Omega^{-1}\,\hat{\nabla}_{[i}\,\hat{L}_{j]k} = 
\Omega^{-4}\,(\hat{\rho} + w)\left(
 U_{k}\,U_{[i}\,\Sigma_{j]}
 - g_{k [i }\,U_{j]}\,\Sigma_{0}
\right)
\end{equation}
\[
+ 4\,\rho\,\left( (1 + w')\,
\Sigma_{[i}\,U_{j]}\,U_{k}
+ \frac{1}{3}\,\Sigma_{[i}\,g_{j]k}\right)
\]
\[
+ \Omega^{-3}\,
(\hat{\rho} + w)
\left(\nabla_{[i}\,U_{j]}\,U_{k}  
+ U_{[j}\,\nabla_{i]}\,U_{k}  
\right)
\]
\[
+ \Omega\,\left((1 + w')\,
\nabla_{[i}\,\rho\,\,U_{j]}\,U_{k}  
+ \frac{1}{3}\,\nabla_{[i}\,\rho\,g_{j]k}\right).
\]
Its components, given by 
\begin{equation}
\label{comp-dust-0b0-nabla-hat-L-hat}
M_{0b0} = 
\Omega^{-4}\,
 \frac{\hat{\rho} + w}{2}\,(\Sigma_{b} - \Omega\,f_b)
 - \frac{2 + 3\,w'}{6}\,(\Omega\,\nabla_{b}\,\rho
+ 4\,\rho\,\Sigma_b),
\quad
\end{equation}
\begin{equation}
\label{ON-dust-ab0-nabla-hat-L-hat}
M_{ab0}  = - \Omega^{-3}\,
(\hat{\rho} + w)
\,\chi_{[ab]}
\quad \quad \quad \quad \quad \quad  \quad \quad \quad 
\quad \quad \quad \quad 
\quad \quad \,
\end{equation}
\begin{equation}
\label{comp-dust-0bc-nabla-hat-L-hat}
M_{0bc} = 
\left\{- 
\Omega^{-4}\,\frac{\hat{\rho} + w}{2}\,\Sigma_{0}
+ \frac{1}{6}\,(\Omega\,\nabla_0\rho + 4\,\rho\,\Sigma_0)
\right\} g_{bc}
\quad \quad \quad \quad \,\,\,    
\end{equation}
\begin{equation}
\label{comp-dust-abc-f-nabla-hat-L-hat}
M_{abc} =  
\frac{1}{3} \left\{\Omega\,\nabla_{[a}\,\rho\,g_{b]c}
+ 4\,\rho\,\Sigma_{[a}\,g_{b]c}\right\},
\quad \quad \quad \quad \quad \quad  \quad \quad \quad 
\quad \,\,\, 
\end{equation}
are regular if $4\,k \in \mathbb{N}_0$.

\section{Cauchy problems.}

To solve the conformal equations with an asymptotic radiation equation of state we extract from the complete system a symmetric hyperbolic  {\it reduced system} for the unknown
\[
{\bf Z} = (\Omega, \quad \Sigma_k, \quad s, \quad e^{\mu}\,_a,  \quad \Gamma_a\,^b\,_c, \quad \rho, \quad \chi_{ab}, \quad f_a, \quad L_{jk}, \quad C^i\,_{jkl}),
\]
where the Ricci scalar $R = R[g] = 6\,L_j\,^j$
of $g_{\mu\nu}$ plays the role of a conformal gauge source function that controls implicitly the evolution of the conformal factor. It can be prescribed  as an arbitrary function of the coordinates.
The present discussion is concerned with the transition from the `physical part of the solution' to scri and beyond.
In physical terms this involves a discussion of a domain of infinite temporal extent but in conformal terms it involves only a finite step in the conformal time $\tau$. The specification of $R[g]$ is thus not particularly  critical. 
{\it It is  chosen to be constant}. 

\vspace{.2cm}

\noindent
{\bf Symmetric hyperbolic reduced equations.}

\vspace{.2cm}

The reduced system is essentially built from the equations which contain derivatives in the direction of $U$. However, sometimes these equations are modified by using constraints, i.e. equations which do not contain derivatives in the direction of $U$.
For the first six components of {\bf Z} we get the system
\[
e_0(\Omega) = \Sigma_0,
\]
\[
e_0(\Sigma_0) - f^c\,\Sigma_c
= - \Omega\,L_{00} - s
+ \Omega^{-1}\,\frac{3\,(\hat{\rho} + w)}{8},
\]
\[
e_0(\Sigma_a) - f_a\,\Sigma_0 = - \Omega\,L_{0a}, 
\]
\[
\nabla_{0}\,s + \Sigma^{i}\,L_{i 0} = 
\frac{1}{8}\,\Omega^{-2}\left\{(1 - 3\,w')\,w - 3\,(\hat{\rho} + w)\right\}
\Sigma_0
\]
\[
- \frac{1}{24}\,(1 - 3\,w')\,
 (\Omega^{3}\,\rho + \Omega^{-1}\,w)\,\chi_{a}\,^{a},
\]
\hspace*{2.5cm}where (\ref{E-conf-w-energy}) 
resp. (\ref{rad-E-conf-w-energy}) 
has been used to replace 
$\nabla_0\rho$,
\[
e^{\mu}\,_{a,\,0}
= - \chi_{a}\,^{b}\,e^{\mu}\,_{b}
+  f_{a}\,\delta^{\mu}\,_{0},
\]
\[
\Gamma_a\,^b\,_{c,\,0} =
- f^b\,\chi_{ac} + \chi_a\,^b\,f_c
- \chi_{a}\,^d\,\Gamma_d\,^b\,_c
+ \Omega\,W^b\,_{c0a}
- g^b\,_{a}\,L_{0 c} + L^b\,_{ 0}\,g_{a c},
\]

\[
\nabla_{0}\rho
+ (\rho + \Omega^{-4}\,w)\,\chi_{a}\,^{a}
+ \Omega^{-1}\,(\rho - 3\,\Omega^{-4}\,w)\,\Sigma_{0} = 0,
\]

\vspace{.1cm}

\noindent
Adding a suitable contraction of (\ref{s0b-Gamma-constr})
to equation (\ref{0a-conf-2nd-U-rho-int-cond}) gives
the equation
\begin{equation}
\label{fa-sym-hyp}
e_0(f_a)  - w'\,e_c(\chi_a\,^c) 
= - \chi_{ac}\,f^c - (1 - 3\,w')\,L_{a0} 
+ w'\,\chi_c\,^c\,f_a 
\end{equation}
\[
+ (1 - 3\,w')\,(\Omega^{-1}\,\nabla_0\Omega\,f_a 
+ \Omega^{-1}\,\chi_{ab}\,\nabla^b\Omega  
- \Omega^{-2}\,\nabla_0\Omega\,\nabla_a\Omega)
\]
\[
- w''\,\frac{\hat{\rho} + w}{w'}\,(\chi_c\,^c\,f_a 
- 3\,\Omega^{-1}\,\nabla_0\,\Omega\,f_a - 
\Omega^{-1}\,\chi_c\,^c\,\nabla_a\Omega
+ \Omega^{-2}\,\nabla_0\Omega\,\nabla_a\Omega),
\]
\[
+ w'\,(2\,\chi_{[c}\,^e\,\Gamma_{a] e}\,^c - 2\,f^c\,\chi_{ca} 
- 2\,\Gamma_{[c}\,^e\,_{a]}\,\chi_e\,^c - 2\,L^0\,_{a}).
\]
Adding (\ref{ab-conf-2nd-U-rho-int-cond}) to
(\ref{s2b-Gamma-evol}) gives
\begin{equation}
\label{chiab-sym-hyp}
e_0(\chi_{ba}) - e_a(f_{b}) 
=
- f_d\,\Gamma_a\,^d\,_b
+ f_a\,f_b 
- \chi_a\,^d\,\chi_{db}
+ \Omega\,W^0\,_{b0a}
+ L_{a b} + L^0\,_{ 0}\,g_{a b},
\end{equation}
\[
- 2\,(\chi_{c[a}\,\chi_{b]}\,^c  
+ w'\,\chi_c\,^c\,\chi_{[ab]}
+ (1 - 3\,w')\,\Omega^{-1}\,\Sigma_0\,\chi_{[ab]}).
\]
If equations (\ref{fa-sym-hyp}) and (\ref{chiab-sym-hyp})
are written as a system for $f_a$ and $\chi_{bc}$ with principal part 
\[
\frac{1}{w'}\,e_0(f_a) - e_c(\chi_a\,^c) = \ldots\,,
\quad \quad
e_0(\chi_{ba}) - e_a(f_b) = \ldots \,.
\]
 they represent, with given right hand sides, a symmetric hyperbolic system.

\vspace{.2cm}

To derive the reduced systems for the curvature fields we follow the methods of \cite{friedrich:dust:2016} and earlier articles. 
Equation (\ref{rad-f-conf-G-coord-L-equ}) provides a system for the unknowns $L_{0a} = L_{a0}$ and $L_{ab} = L_{ba}$  that  
contains no derivatives of $L_{00}$. Where this quantity appears in the equation the contraction $- L_{00} + g^{ab}\,L_{ab} = L_j\,^j = R/6$ can be used to express it in terms of $L_{ab}$ and 
the gauge source function $R = R[g] = const$.
The system is given by

\begin{equation}
\label{L0a-evol}
 \nabla_0\,L_{0a}  - g^{bc}\,\nabla_b\,L_{a c} = 
\Omega\,(M_{0a0} - g^{bc}\,M_{bac})
, \quad a = 1, 2, 3,
\end{equation}
\begin{equation}
\label{L11-etc-evol}
\nabla_{0}\,L_{a a} - \nabla_{a}\,L_{0 a} = 
\Sigma_{l}\,W^{l}\,_{a 0 a} + \Omega\,M_{0aa}, \quad \,\,\,
\quad a = 1, 2, 3,
\end{equation}
\begin{equation}
\label{L12-etrc-evol}
2\,\nabla_{0}\,L_{a b} 
- \nabla_{a}\,L_{0 b} - \nabla_{b}\,L_{0 a}  = 
- 2\,\Sigma_{l}\,W^{l}\,_{(ab)0}
+ 2\,\Omega\,M_{0(ab)},\,\,\,
\end{equation}
\[
a,  b  = 1, 2, 3, \,\,a < b.
\]
To extract the desired system from (\ref{rad-f-conf-G-coord-W-equ}) 
consider the fields
\[
h^j\,_k = g^j\,_k +U^j\,U_k, \,\,\, l^j\,_k  = g^j\,_k +2\,U^j\,U_k,
\,\,\,
 \epsilon_{ijkl} =  \epsilon_{[ijkl]}, 
 \,\,\, \epsilon_{jkl} = U^i\, \epsilon_{ijkl},
 \,\,\, \mbox{with} \,\,\, \epsilon_{0123} = 1.
 \]   
The symmetric, trace-free $U$-electric part $w_{jl}$ and the $U$-magnetic part $w^*_{jl}$ of $W^{i}\,_{j k l}$, 
\[
w_{jl} =   W_{i p k q}\,U^i \,h^p\,_j\,U^k\,h^q\,_l, \quad \quad 
w^*_{jl} =   1/2\,W_{i p m n}\,\epsilon^{mn}\,_{kq}\,U^i \,h^p\,_j\,U^k\,h^q\,_l, 
\]
allow us to represent the rescaled conformal Weyl tensor in form 
\[
W_{ijkl} = 2\,(l_{i[k}\,w_{l]j} - l_{j[k}\,w_{l]i} 
- U_{[k}\,w^*_{l] p}\,\epsilon^p\,_{ij}
- U_{[i}\,w^*_{j]p}\,\epsilon^p\,_{kl}).
\]
Suitable evolution equations for $w_{jl}$ and $w^*_{jl}$ 
is given by 
$\nabla^i\,W_{i(a|dc|}\,\epsilon_{b)}\,^{dc} = 
K_{dc(a}\,\epsilon_{b)}\,^{dc}$, 
$\nabla^i\,W_{i(a|0|b)} = K_{0(ab)}$
which read in detail
\begin{equation}
\label{a-rad-w*-evol}
e_0\,(w^*_{ab}) - D_d\,w_{c(a}\,\epsilon_{b)}\,^{dc} =
- \chi_{e d}\,w^*_{c f}\,\epsilon_{(a}\,^{f e}\,\epsilon_{b)}\,^{dc} 
\quad \quad \quad \quad \quad \quad \quad \quad \quad
\end{equation}
\[
\quad \quad \quad \quad 
+ (\chi^e\,_{(a}\,w^*_{b)e} - \chi\,w^*_{ab})
+  f_{d}\,w_{c(a}\,\epsilon_{b)}\,^{dc},
\]
and 
\begin{equation}
\label{a-rad-w-evol}
e_0 (w_{ab}) 
+ D_d\,w^*_{c (a}\,\epsilon_{b)}\,^{dc}
= - 2\,\chi\,w_{ab} 
+ 2\,\chi^c\,_{(a}\,w_{b) c}
\quad \quad \quad \quad \quad \quad \quad \quad 
\end{equation}
\[
\quad \quad \quad \quad 
+ \chi_{(a}\,^c\,w_{b)c} 
- \chi^{cd}\,w_{cd}\,\,g_{ab} 
+ 2\,w^*_{(a}\,^e\,\epsilon_{b)ed}\,f^d
 - M_{0(ab)},
\]
with `spatial derivatives' 
$D_a\,w^*_{bc} \equiv e_a\,(w^*_{bc}) 
- \Gamma_a\,^e\,_b\,w^*_{ec} - \Gamma_a\,^e\,_c\,w^*_{be}
= \nabla_a\,w^*_{bc}$.
 The components of $M_{jkl}$ on the right hand sides of equations
(\ref{L0a-evol}) -  (\ref{a-rad-w-evol})
are given by
\[
M_{0a0} - g^{bc}\,M_{bac} = 
\Omega^{-4}\,
 \frac{\hat{\rho} + w}{2}\,(\Sigma_{a} - \Omega\,f_a)
 - \frac{w'}{2}\,(\Omega\,\nabla_{a}\,\rho
+ 4\,\rho\,\Sigma_a),
\]
\[
M_{0ab} = - \left\{ 
\Omega^{-4}\,\frac{\hat{\rho} + w}{2}\,\Sigma_{0}
- \frac{1}{6}\,(\Omega\,\nabla_0\rho + 4\,\rho\,\Sigma_0)
\right\} g_{ab}, \quad \quad 
M_{dc(a}\,\epsilon_{b)}\,^{dc} = 0,
\]
where here and in (\ref{L0a-evol}), (\ref{L11-etc-evol}), (\ref{L12-etrc-evol})
equations (\ref{rad-E-conf-w-energy}) and 
(\ref{rad-E-conf-w-flow})
must be used to replace $\nabla_{0}\rho$ and $\nabla_a\rho$.

\vspace{.1cm}

\noindent
Observing the symmetry of $w_{ab}$,  $w^*_{ab}$  (but ignoring their trace-freeness, which can later be recovered as a consequence of the initial data and the equations above), equations (\ref{a-rad-w*-evol}) and 
(\ref{a-rad-w-evol}) can written as system for the unknowns
$w_{ab}$ and $w^*_{ab}$ with $a \le b$, $a, b = 1, 2, 3$.
If the equations $w_{ab,\,0} = \ldots$, $w^*_{ab,\,0} = \ldots$
are then written with a factor 2 if $a < b$, the system is seen to be manifestly symmetric hyperbolic. 

\vspace{.1cm}

\noindent
Together with the previous equations we have obtained now a 
quasi-linear system for the unknown {\bf Z} which can be written in the form
${\bf A}^{\mu}\,\partial_{\mu}{\bf Z} = {\bf B}$
with matrix-valued functions ${\bf B}  = {\bf B}({\bf Z})$ and  
${\bf A}^{\mu} = {\bf A}^{\mu}({\bf Z})$ that 
are symmetric, i.e. $^t{\bf A}^{\mu} = {\bf A}^{\mu}$,  with
${\bf A}^{0}$ defining a positive definite bilinear form if the $e^0\,_a$
are not too large.

\vspace{.3cm}

\noindent
{\bf Cauchy data for the conformal equations.}

\vspace{.3cm}

Solutions that admit smooth conformal extensions at future time-like infinity can be constructed from data for the conformal field equations on a compact Cauchy hypersurface $S$ in the `physical domain', where $\Omega > 0$, or on the compact 3-manifold $S = {\cal J}^+ = \{\Omega = 0\}$ that represents  future time-like infinity.

Unless the field $U$ is assumed to be orthogonal to $S$,
the data, which must satisfy the constraints induced by the conformal field equations on $S$,  are in general expressed in terms of the unit normal to  $S$ and then transformed into a frame $e_k$ with $e_0 = U$. This requires calculations involving the complete system of conformal field equations which are fairly tedious (see \cite{friedrich:nagy} where the presence of a boundary requires this) and give little insight. The calculation will be skipped here.

We will first construct 
{\it asymptotic end data} on ${\cal J}^+$ with the assumption that
$U$ is orthogonal to ${\cal J}^+$. It is interesting to note that in the case of a pure dust equation of state one is forced into this requirement without forcing the field $U$ to be hypersurface orthogonal in the physical domain (see \cite{friedrich:dust:2016}). In the present case 
 this assumption implies a genuine restriction. 
The analysis of the asymptotic end data follows essentially the one given in the vacuum case \cite{friedrich:1986a}
 with some modifications if matter fields are involved \cite{friedrich:massive-fields},
\cite{friedrich:dust:2016}

To analyse the constraints, i.e. the equations that do not involve derivatives in the direction of $U$, we assume that $k \ge 1$. The restrictions of  the equations to ${\cal J}^+$
are then regular and the expressions containing the quantities $w^*$, $(w^*)'$ drop out. 
 The unknowns are then determined as follows.

 The requirement that $U$ is orthogonal to ${\cal J}^+$ implies
\[ 
 e^0\,_a = 0, \quad \Sigma_a = 0, \quad
 \chi_{ab} \,\,\,\mbox{is the second fundamental form induced on $ {\cal J}^+ $}.
 \]
The first structural equation (\ref{conf-O-torsion-free condition})
implies on $ {\cal J}^+ $ the first structural equation with respect to the
the frame $e_a$ and metric $h_{ab}$  induced on ${\cal J}^+$.
The $ \Gamma_{a}\,^{c}\,_{b}$ are thus the connection coefficients
of the covariant derivative operator $D$ induced by the metric $h_{ab}$ in the frame $e_a$.
 The second structural equation (\ref{conf-O-Ricci identity}) reduces to the relation
$R^a\,_{cdb}[h] = 2\,g^a\,_{[d}\,L_{b] c} + 2\,L^a\,_{ [d}\,g_{b] c}$,
 equivalent to 
\[
L_{ab}[g] = l_{ab}[h] \quad \mbox{with} \quad 
l_{ab}[h] = R_{ab}[h] - \frac{1}{4}\,R[h]\,h_{ab}.
\]
This gives $L_{00} = g^{ab}\,L_{ab} - L_j\,^j = R[h]/4 - R/6 $
where $R = R[g]$ represents the conformal gauge source function.
Equations (\ref{E-conf-w-flow}), (\ref{f-G-coord-alg-equ}),
(\ref{f-G-coord-Omega-equ}), (\ref{f-G-coord-s-equ}) imply
\[
f_a = - \frac{1}{4}\,\nabla_{a}\rho, \quad
\Sigma^0 = \nu \equiv \sqrt{\lambda/3} > 0, \quad 
\chi_{a b} = s/\nu\,g_{ab}, \quad 
\nabla_{a}\,s =  - \nu\,L_{0a},
\]
where it is assumed that $U$ is future directed and $\Omega$ is decreasing in the direction of $U$. By a rescaling $g_{\mu\nu} \rightarrow \theta^2\,g_{\mu\nu}$ and
$\Omega \rightarrow \theta\,\Omega$, where 
$\theta > 0$ is smooth with prescribed value and suitable normal derivative on ${\cal J}^+$, it can be achieved that 
\[
s = 0 \quad \mbox{whence} \quad \chi_{ab} = 0, \quad L_{0a} = 0 \quad
\quad \mbox{on} \quad {\cal J}^+.
\]   
Equation (\ref{f-conf-G-coord-L-equ}), which reduces to
\[
D_{c}\,l_{d a}\,\epsilon_b\,^{cd} = \nu\,w^*_{ab},
\]
relates the Cotton tensor of $h$, given on the left hand side, to the 
$U$-magnetic part of $W^i\,_{jkl}$.
The constraints induced by (\ref{f-conf-G-coord-W-equ}), given by
$\nabla_{l}\,W^{l}\,_{0 i j} = M_{ij0}$,
translate with $M_{ab0} = 0$ and $M_{0b0} = 0$
to the constraint
\begin{equation}
\label{weyl-constr}
D^a\,w_{ab} = 0,
\end{equation}
and to $D^a\,w^*_{ab} = 0$, which is not a constraint but the identity satisfied by the Cotton tensor. 

\vspace{.1cm}
\noindent
If the data
\[
e^{\alpha}\,_a, \quad   
\rho, \quad  
 w_{ab},
\]
are prescribed on ${\cal J}^+$ so that
 $h^{\alpha\beta} =\delta^{ab}\,e^{\alpha}\,_a\,e^{\beta}\,_b$ defines the contravariant version of a Riemannian metric, $\rho > 0$, and 
 $w_{ab} = w_{ba}$, $w_c\,^c = 0$ with   $w_{ab}$
 satisfying (\ref{weyl-constr})
 with the covariant derivative operator $D$ defined by $h$, all the unknowns subsumed by ${\bf Z}$ can be determined in our gauge.
 
 \vspace{.2cm}
 
 \noindent
 {\bf On the preservation of the constraints.}

\vspace{.2cm}

Smooth data on ${\cal J}^+$ as described above, determine a smooth solution to the symmetric hyperbolic reduced equations that cover a neighbourhood of
${\cal J}^+ = \{\tau = 0\}$ with $\Omega > 0$ in the past of ${\cal J}^+$, where the parameter $\tau$ on the flow lines of $U$ is negative, and 
$\Omega < 0$ in the future of ${\cal J}^+$ where $\tau > 0$
\cite{kato}. We can assume it to exist in a range 
$- \tau_* \le \tau \le \tau_*$ with some $\tau_* > 0$ and $\tau = 0$ on 
${\cal J}^+$.
  
Once this solution has been obtained, it remains to show that the solution to the reduced equations does in fact also satisfy the constraints and thus the complete system of conformal equations.
Since the proof follows a standard recipe which has been worked out in detail in earlier articles (see \cite{friedrich:dust:2016}, and in particular
\cite{friedrich98}, \cite{friedrich:rendall} for a discussion of the Einstein-perfect-fluid equations with a general equation of state), we skip this step.

There remains, however, an open question. Since we assumed 
$U$ to be orthogonal to ${\cal J}^+$ we should expect $U$ to be hypersurface orthogonal, that is $\chi_{[ab]} = 0$ in the range $- \tau_* \le \tau \le \tau_*$. Because $\chi_{[ab]} = 0$ on ${\cal J}^+$,
one could think this to be a consequence of equation 
(\ref{ab-conf-2nd-U-rho-int-cond}). Since this equation is neither a constraint nor satisfied as part of the reduced system, we use a different argument. The evolution of $\chi_{ab}$ is governed by equation (\ref{chiab-sym-hyp}), which implies
\[
e_0(\chi_{[ba]}) = D_{[a}f_{b]} +
 \chi_{[ac]}\,\beta_b\,^c
-  \chi_{[bc]}\,\beta_a\,^c
- 2\,w'\,\chi_c\,^c\,\chi_{[ab]}
- 2\,(1 - 3\,w')\,\Omega^{-1}\,\Sigma_0\,\chi_{[ab]},
\]
with  $\beta_{ab} =  \chi_{(ac)}$.We formally write here
$D_a f_{b} = e_a(f_{b}) - f_d\,\Gamma_a\,^d\,_b$,
not implying that $D$ denotes a covariant derivative away from ${\cal J}^+$.
Among the constraints which we can assume to be satisfied  there is 
the relation (\ref{E-conf-w-flow}) or its more concise version  (\ref{f_a-meaning}), i.e.
\begin{equation}
f_a = e_a(f)
\quad \mbox{with} \quad f = F + \log \Omega
\quad \mbox{and} \quad F' = - (\hat{\rho} + w)^{-1}\,w'.
\end{equation}
Since the connection $\nabla$  is torsion free it follows 
\[
0 = \nabla_a\nabla_b f - \nabla_b\nabla_af =
2\,(D_{[a} f_{b]} - \chi_{[ab]}\,e_0(f)),
\]
where the function 
\[
e_0(f) = e_0(F + \log \Omega) = - \frac{w'}{\rho + \Omega^{-4}\,w}\,\nabla_0\rho
+ \Omega^{-1}\,\frac{\rho + \Omega^{-4}\,w - 4\,\rho\,w'}
{\rho + \Omega^{-4}\,w}\,\nabla_0\Omega,
\]
is regular if $k \ge 1$. 
The resulting ODE for $\chi_{[ab]}$ implies that $\chi_{[ab]} = 0$,
equation (\ref{ab-conf-2nd-U-rho-int-cond}) is satisfied,
and $U$ is hypersurface orthogonal in the range $- \tau_* \le \tau \le \tau_*$.

\vspace{.2cm}

With this we discussed all the ingredients needed to obtain existence results. General results in symmetric hyperbolic systems
\cite{kato}
allow us to draw the following conclusions.

\vspace{.2cm}

\noindent
{\it Let the reduced conformal Einstein-$\lambda$-perfect-fluid equations with asymptotic radiation equation of state where 
$k \ge 1$ 
be given in terms of the gauge discussed above so that
$e_0 = U$.
Assume the future directed flow vector $U$ to be
orthogonal to the compact 3-manifold ${\cal J}^+ = \{\Omega = 0\}$. 
Let on ${\cal J}^+$ be given a minimal end data set
consisting of a smooth frame field $e_a = e^{\alpha}\,_a\,\partial_{x^{\alpha}}$, a positive function $\rho$, and a symmetric trace-free tensor field $w_{ab}$ that satisfies $D^a\,w_{ab} = 0$
where $D$ is the covariant derivative operator associated with the metric $h$ that satisfies $h(e_a, e_b) = \delta_{ab}$. Then:}

\vspace{.1cm}

\noindent
$-$ {\it As discussed above, a complete set of Cauchy data for the conformal field equations
can be calculated from the minimal set if the gauge conditions, the constraints, and the special features of ${\cal J}^+ = \{\Omega = 0\}$ are taken into account.} 

\vspace{.1cm}

\noindent
$-$ {\it These data determine a smooth solution $\Omega, \,g, \, \ldots,\,$ to the conformal field equation 
(unique up to extensions) so that $\Omega < 0$ in the future of 
${\cal J}^+$,  $\Omega > 0$  in the past of ${\cal J}^+$, and 
 $\hat{g} = \Omega^{-2}\,g$, $\hat{U} = \Omega\,U$, 
$\hat{\rho} = \Omega^4\,\rho$ satisfy where $\Omega \neq 0$
the
Einstein-$\lambda$-perfect-fluid equations  with asymptotic radiation equation of state. The flow field $\hat{U}$ is hypersurface orthogonal.}

\vspace{.1cm}

\noindent
$-$ {\it Let $S$ by a Cauchy hypersurface in the `physical domain' of this solution where $\Omega > 0$ and denote by $\Delta$ the Cauchy data induced by the solution on $S$. Let $\Delta'$ be Cauchy data
on $S$ for the same system of equations. If these data are sufficiently close to $\Delta$ they develop into a solution which is time-like and null geodesically future complete, admits a smooth conformal boundary 
${\cal J}'^{+}$ representing its future time-like infinity and a smooth conformal extension beyond with $\rho > 0$.}

\vspace{.1cm}

\noindent
$-$ {\it If the cases where $w^* = 0$ is admitted, the resulting set of solutions contains as special cases the FLRM-pure-radition solutions which develop a smooth conformal boundary in the future}. 

\vspace{.2cm}

Given the data $\Delta'$ whence the corresponding flow field $U'$ on $S$, the field equations allow us to determine the quantity 
$U'_{[i}\,\nabla'_j U'_{k]}$ on $S$. In the last statement it is not required that this quantity vanishes on $S$. Thus the flow field $U'$ determined by the data $\Delta'$ will not necessarily be hypersurface orthogonal and $U'$ need not be orthogonal to ${\cal J}'^{+}$.

\vspace{.2cm}

If Cauchy data are given on ${\cal J}^+$ with a flow vector field $U$ that is not necessarily orthogonal to ${\cal J}^+$ the result stated at the end of the introduction is obtained.

}

\end{document}